\documentclass{nature}
\usepackage{amsthm,amsmath,amssymb,authblk,tikz,graphicx}
\usepackage[labelfont={small,bf},textfont={footnotesize,it}]{caption}
\usepackage{subcaption,algorithm2e}
\usepackage{listings}
\newcommand*\rfrac[2]{{}^{#1}\!/_{#2}}

\usepackage[utf8]{inputenc}

\newlength\tindent
\usetikzlibrary{shapes,decorations,circuits.logic.US,circuits.logic.IEC,fit,external}
\tikzstyle{loosely dashed}=[dash pattern=on 4pt off 8pt]
\tikzstyle{loosely dashed2}=[dash pattern=on 6pt off 8pt]

\title{The physics of (ir)rational choice}

\author{Joost Kruis$^{1\ast}$, Gunter Maris$^2$, Maarten Marsman$^1$, Dylan Molenaar$^1$, Maria Bolsinova$^2$ \& Han L. J. van der Maas$^1$}

\begin{document}

\maketitle

\begin{affiliations}
 \item University of Amsterdam
 \item ACT-Next by ACT
\end{affiliations}

\begin{abstract}
Even though classic theories and models of discrete choice pose man as a rational being, it has been shown extensively that people persistently violate rationality in their actual choices. Recent models of decision-making take these violations often (partially) into account, however, a unified framework has not been established. Here we propose such a framework, inspired by the Ising model from statistical physics, and show that representing choice problems as a graph, together with a simple choice process, allows us to explain both rational decisions as well as violations of rationality.
\end{abstract}

\section*{(Ir)rational choice}

Over the last century, our perception of peoples choice behaviour as being rational, consistent, and profit-maximising, shifted to one where we now describe human decision-making as biased, uncertain, and settling for a satisfactory outcome. The feature thought to determine the appeal of choice alternatives, first changed from expected value\cite{huygens1657,laplace1814}, to expected utility\cite{bernoulli1954,neumann1955}, and then to the prospect of the alternative\cite{kahneman1979}. Instead of mulling over a decision until it was clear what the best option was, people where shown to stop when an acceptable alternative had been found, a consequence of what was termed bounded rationality\cite{simon1955}. Rather than an unwavering regularity in their choices, people behave rather inconsistently, for example because of the endowment effect, which entails that people tend to value an alternative higher as soon as they have chosen it, compared to when they have not\cite{thaler1980}.  

Two competing classes of decision-making models are random- and fixed utility models, the first assuming that choice is deterministic but with the appeal of the alternatives as a random variable that changes over time\cite{thurstone1927, marschak1960, mcfadden1974}, whereas the second posits that the appeal of alternatives is fixed, but the decision process is of probabilistic nature\cite{luce1959,luce1965}. These models of choice often assume choices to be rational in nature by endowing them with properties such as the independence of an alternative's appeal from irrelevant alternatives, and the existence of a, either weak or strong, transitivity in the ordering of choice appeal\cite{thurstone1927,luce1959}. Although these are statistically desirable properties, it has been shown that they are often violated. Most well known of these violations are the similarity, attraction and compromise effect\cite{debreu1960,restle1961,becker1963,luce1965,krantz1967,rumelhart1971,tversky1972,luce1977,simonson1989,simonson1992,tversky1993}. 

With the increased understanding of the complexity of human decision-making, both the theories and models describing choice behaviour have become increasingly complex as well\cite{tversky1972, roe2001}. Although these models take violations of rationality into account, this is either the case for one of the violations and not the other, or only for violations taking place under specific conditions. Developing a framework that integrates these (partial) theories and explains the key choice phenomena is thus a worthwhile effort. 

In this article we propose a simple choice framework, that allows us to explain both rational and irrational choices. We achieve this by framing choices as a combination of cues and alternatives and the relation between them, and introduce a simple decision process that stops when the choice conditions are met for the first time, as such implementing bounded rationality. Using only this decision rule, known violations of rationality, and other choice phenomena such as the endowment effect, follow naturally from the model. 

\section*{Choice Setup}

Choices are framed as combinations of cues and alternatives. Cues represent the conditions of the choice, e.g. `buy a book' or `select a present', and alternatives describe the possible choices. The relations between these alternatives and cues, and between alternatives themselves, are what govern the initial conditions of a choice. An appropriate representation of this framework is a graph, the vertices of which correspond to alternatives and cues, and the edges describing the relationship between the vertices. Figure \ref{fig:01} shows two examples of possible choice graphs.

\begin{figure}
    \centering
    \begin{tikzpicture}[every node/.style = {draw=none, text=black, circle, minimum size = 13mm, fill=gray!25}]
        \path
        (8,3) node(y2) {$C$}
        (5,0) node[draw, line width=5pt](x21) {$A_1$}
        (8,0) node[draw, line width=3pt](x22) {$A_2$}
        (11,0) node[draw, line width=.5pt,dashed](x23) {$A_3$};
        \draw [line width=1pt,-,black] (y2) -- (x21);
        \draw [line width=2pt,-,black] (y2) -- (x22);
        \draw [line width=3pt,-,black] (y2) -- (x23);
        \draw [line width=2pt,-,loosely dashed,black] (x22.south east) to[bend right=40] (x23.south west);
    \end{tikzpicture}
    \caption{\textbf{Graph for a choice with one cue ($C$), and three alternatives ($A_1, A_2, A_3$).} \\ Solid edges represent positive relations between vertices $(\omega > 0)$, whereas dashed edges represent negative relations $(\omega < 0)$. Base appeal of the alternatives $(\mu)$ is represented by the ring around the vertex. Solid rings represent appealing alternatives $(\mu>0)$, whereas a dashed rings represent unappealing alternatives $(\mu<0)$. The thickness of both the edges and rings around the vertices corresponds to intensity of the relation/appeal.}
    \label{fig:01}
\end{figure}

To formalise a choice graph with $n$ vertices as a probability distribution, we let $\boldsymbol{\sigma} = [\sigma_1, \sigma_2, \dots, \sigma_n]$ be the state vector of the $n$ vertices, where $\sigma_i \in \{0,1\}$ denotes if vertex $i$ is active $(\sigma_i = 1)$ or inactive $(\sigma_i = 0)$, and endow the states with the following distribution;

\begin{equation}
    \boldsymbol{\sigma} \sim p(\boldsymbol{\sigma}) = \frac{1}{Z_{\beta}} \exp{\left( \beta \left[\sum_i \sigma_i \, \mu_i + \sum_{\langle i,j \rangle} \sigma_i \sigma_j \, \omega_{ij} \right]\right)} \, .
    \label{eq:01}
\end{equation}

Equation \ref{eq:01} can be recognised as an Ising model\cite{lenz1920,ising1925} originally introduced in statistical physics to model the ferromagnetic behaviour of magnetic dipole moments of atomic spins on a lattice. Where $\mu_i$ is the force of the external magnetic field interacting with spin $i$, $\omega_{ij}$ the pairwise interaction between spins $i$ and $j$, and $\beta$ the inverse temperature of the system. $Z_{\beta}$ is the appropriate, temperature dependent, normalising constant to make the probabilities sum to one, also known as the partition function. While this application was later rejected, due to incompatibility with quantum mechanics, in favour of the Heisenberg model, it has remained highly popular and one of the most studied models in modern statistical physics\cite{niss2005}. Also outside the field of physics the Ising model has attracted attention as a way to model the joint distribution of binary variables as a function of main effects and pairwise interactions\cite{marsman2015,epskamp2016,kruis2016,marsman2018}. 

In the current context $\mu_i$ can be perceived as the base appeal of cue/alternative $i$, whereas $\omega_{ij}$ specifies the strength of the relation between cues and/or alternatives $i$ and $j$. In Figure \ref{fig:01} it is shown how differences in $\mu$ and $\omega$ are represented in the choice graphs. In equation \ref{eq:01} can be seen that the inverse temperature $\beta$ scales the magnitude of both the base appeal and pairwise interactions. While the quantity of $\beta$ is known in thermodynamics as a function of the temperature and the Boltzmann constant, in the current application it is not identifiable. It can however be interpreted, in the context of decision-making, as a measure of choice consistency, as with lower $\beta$ choice behaviour becomes increasingly random, until $\beta = 0$, where the choice is purely random. It is assumed that before a person is faced with a choice, the internal state of the decision maker, the resting state, is in accordance with the Ising model from equation \ref{eq:01}. 

When a person is confronted with a choice, the appropriate cue vertex $(c)$ gets activated $(\sigma_c = 1)$, and remains active during the choice process. Second, to represent that a single alternative needs to be chosen, lateral inhibition, a neural concept where an exited neuron inhibits its direct neighbours, is introduced between the alternatives that are part of the choice\cite{feldman1982}. In the framework this is represented by lowering the interaction between each pair of alternatives with a constant $(\varepsilon)$, such that otherwise independent alternatives become negatively related. The consequence of this inhibition is that a choice structure will naturally tend to move to a state in which, apart from the cue vertex, only one alternative will remain activated\cite{yuille1989}. Furthermore, the stronger the lateral inhibition, the faster the system will tend to the state with only one alternative remaining active. Factors such as time-pressure, importance of a choice (e.g. buying a house vs. buying a book), and general (in)decisiveness might be possible contributors to variations in both $\beta$ and $\varepsilon$. Imposing the condition that, next to the cue vertex, only one additional alternative vertex can remain active $(\sum_{i=1}^n \sigma_i = \sigma_+ = 2)$, on the resting state distribution, represent the choice condition where a single alternative needs to be chosen. Enforcing these conditions on the Ising model we obtain conditional choice probabilities for each alternative.

\begin{align}
    p(\sigma_{a\neq c} = 1 \mid \sigma_c = 1,\sum_{i=1}^n \sigma_i = \sigma_+ = 2) &= \frac{\exp{\left(\beta\left[\mu_a +  \, \omega_{ca}\right]\right)}}{\sum_{j \neq c} \exp{\left(\beta\left[\mu_j +  \, \omega_{cj}\right]\right)}} \, .
     \label{eq:02}
\end{align}

Equation \ref{eq:02} can be recognised as a multinomial logistic regression model, and shows that under the imposed conditions, the pairwise interactions between alternatives, either from pre-existing conditions or lateral inhibition, vanish, and the probability of choosing alternative $a$ is only dependent on its base appeal ($\mu_a$) and relation with the cue ($\omega_{ca}$). The conditional distribution thus preserves both independence of irrelevant alternatives, and stochastic transitivity, that is, the model has rational decision-making as its ground state.  

\section*{Choice Process}

Why irrational choice behaviour does occur is explained by the choice process that describes how a person moves from an instance of the Ising model in equation \ref{eq:01}, to an instance of the appropriately conditioned Ising model as seen in equation \ref{eq:02}. We model the choice process, one of sequential elimination of alternatives, as a Markov process with the conditioned choice probabilities as their unique invariant distribution. For expository purposes we consider one of the simplest examples of such a process, a Metropolis algorithm with single-spin-flip dynamics\cite{newman1999}.

Let $\boldsymbol{\sigma}$ denote the current state of the choice structure, then at every step in the decision process one alternative is randomly selected $(\sigma_s)$ and its value flipped $(\sigma^*_s = (\sigma_s-1)^2)$. The resulting configuration is our proposal state, denoted by $\boldsymbol{\sigma}^*$, which we accept with the appropriate probability. 

\begin{align}
    p(\boldsymbol{\sigma} \rightarrow \boldsymbol{\sigma}^*) &= \min\left(1,\frac{\exp{\left( \beta \left[\sum_i \sigma^*_i \, \mu_i + \sum_{\langle i,j \rangle} \sigma^*_i \sigma^*_j \, \omega^{\varepsilon}_{ij} \right]\right)}}{\exp{\left( \beta \left[\sum_i \sigma_i \, \mu_i + \sum_{\langle i,j \rangle} \sigma_i \sigma_j \, \omega^{\varepsilon}_{ij} \right]\right)}}\right) \, ,
     \label{eq:03}
\end{align}

where $\boldsymbol{\Omega}^{\varepsilon}$ is the lateral inhibition adjusted pairwise interaction matrix, in  which the relations among the alternatives are lowered by a constant $(\varepsilon)$. Using these acceptance probabilities ensures that in the long run states are in accordance with the conditioned Ising model. 

From the choice probabilities in equation \ref{eq:02} it is clear that if we let the Markov chain run until convergence, the effect of lateral inhibition and any other (in)direct interactions between choice alternatives will have worn out, and only the direct interactions between cues and alternatives matter. As such, in this case it is guaranteed that choices are rational. However, if a person stops when the conditions (i.e., $\sigma_c = 1$ and $\sigma_+ = 2$) hold for the first time, the same is not true. 

By introducing the rule that a person stops whenever the conditions of the choice problem are met for the first time, the model implements the idea of bounded rationality, which states that, as a function of limited cognitive capacity and time, people will search for an alternative that is satisfactory, instead of the most appealing one\cite{simon1955}. The major contribution of this mechanism is that it predict the occurrence of irrational choices, as we explain next. 

\section*{Results}

For each of the three classic violations of rationality, known as the similarity-, attraction-, and compromise-effect, we provide a simple example and illustrate how the proposed choice structure allows us to explain the observed deviations from rational behaviour as a function of interactions between the alternatives, and the rule that a choice will be made as soon as the choice conditions hold for the first time. As the decision is the result of a process that has Markov properties, we can, as we show in the methods section, for small $n$, analytically derive the expected response frequencies, by establishing a stochastic matrix containing the transition probabilities $p(\boldsymbol{\sigma} \rightarrow \boldsymbol{\sigma}^*)$ for all $2^n$ possible configurations of the system. 

It is important to realise that for each of the upcoming examples more than one possible way exists in which a choice can be structured, such that we obtain expected response frequencies that align with empirical observations. One reason for this is the interchangeability in, for example, the relations between the inverse temperature ($\beta$), and the main effects ($\boldsymbol{\mu}$) and pairwise interactions $(\boldsymbol{\Omega})$, or the relation between the cue and and alternative ($\omega_{ca}$) and the base appeal of the alternative $(\mu_a)$. Furthermore, as some readers may have noticed, while traditionally the Ising model $\sigma$ takes values in $\{-1,1\}$, we use $\sigma \in \{0,1\}$ in the current paper. While there exists a linear transformation of the parameters such that they are mathematically indistinguishable, and we would thus obtain the same response probabilities, their interpretation in a choice context would be radically different. 

We do therefore not propose that the choice structures we present in the next part are to be taken as our theory of the particular decision. Rather, our goal is to demonstrate the utility of the proposed modelling framework, that derived from the Ising model, is capable of predicting both rational decision-making behaviour, and deviations of this rationality. 

\subsection*{Similarity}

The classic example for irrational choice behaviour known as the similarity effect, was given by Debreu\cite{debreu1960}, and goes as follows;
\begin{quote}
``Let the set $U$ have the following three elements:
\begin{itemize}
    \item $D_C$, a recording of the Debussy quartet by the $C$ quartet,
    \item $B_F$, a recording of the eighth symphony of Beethoven by the $B$ orchestra conducted by $F$,
    \item $B_K$, a recording of the eighth symphony of Beethoven by the $B$ orchestra conducted by $K$,
\end{itemize}
The subject will be presented with a subset of $U$, will be asked to choose an element in that subset, and will listen to the recording he has chosen. When presented with $\{D_C, B_F\}$ he chooses $D_C$ with probability $\rfrac{3}{5}$. When presented with $\{B_F, B_K\}$ he chooses $B_F$ with probability $\rfrac{1}{2}$. When presented with $\{D_C, B_K\}$ he chooses $D_C$ with probability $\rfrac{3}{5}$. What happens if he is presented with $\{D_C, B_F, B_K\}$? According to \textit{rationality} (original text: the axiom), he must choose $D_C$ with probability $\rfrac{3}{7}$. Thus if he can choose between $D_C$ and $B_F$, he would rather have Debussy. However if he can choose between $D_C$, $B_F$, and $B_K$, while being indifferent between $B_F$ and $B_K$, he would rather have Beethoven.''
\end{quote}

It is immediately clear that the assumptions underlying rational decision-making, would for the this scenario result in expected choice probabilities that are counter-intuitive. Specifically, in this case one would expect that when presented with $\{D_C, B_F, B_K\}$, $D_C$ would be chosen with probability $\rfrac{3}{5}$, and the remaining $\rfrac{2}{5}$ would be split evenly among $B_F$ and $B_K$. An intuition that has since been proven right in comparable experiments\cite{restle1961,rumelhart1971,tversky1972}. 

One choice structure\footnote{The corresponding parameter values $\{\boldsymbol{\mu}, \boldsymbol{\Omega}, \beta, \varepsilon\}$ are provided in the methods section, together with the computer code that allows for the calculation of the expected response probabilities. However, as these examples are for illustrative purposes, we will not discuss the exact values in the current example, nor the subsequent examples.} that explains the occurrence of the similarity effect, does this by introducing a strong negative association between the two Beethoven recordings, as in shown in Figure \ref{fig:02}. The negative relation between $B_F$ and $B_K$ has no influence on choice probabilities for any of the possible two-element subsets, as such the slightly larger base appeal of $D_C$ will results in choosing $D_C$ with probability $\rfrac{3}{5}$, when presented with $\{D_C, B_F\}$ or $\{D_C, B_K\}$. While still present when presented with $\{B_F, B_K\}$, the negative relation works equally in both ways, such that one chooses $B_F$ and $B_K$ both with probability $\rfrac{1}{2}$. While the conditional distribution of the model from equation \ref{eq:02} predicts that $D_C$ will be chosen with probability $\rfrac{3}{7}$, when a choice has to be made from all three alternatives together, the rule that one stops as soon as the choice conditions hold for the first time, will actually predict that when presented with $\{D_C, B_F, B_K\}$, $D_C$ is chosen with probability $\rfrac{3}{5}$, whereas $B_F$ and $B_K$ are both chosen with probability $\rfrac{1}{5}$. Hence, the explanation of `irrational' (yet intuitive) behaviour rest on the presence of a negative relation between the Beethoven recordings and the stopping rule.

\begin{figure}[ht]
    \centering
    \begin{tikzpicture}[every node/.style = {draw=none, text=black, circle, minimum size = 13mm, fill=gray!25}]
        \path
        (0,3) node(y) {$R$}
        (-3,0) node[draw, line width=3pt](x1) {$B_F$}
        (0,0) node[draw, line width=3pt](x2) {$B_K$}
        (3,0) node[draw, line width=3.5pt](x3) {$D_C$};
        \draw [line width=1pt,-,black] (y) -- (x1);
        \draw [line width=1pt,-,black] (y) -- (x2);
        \draw [line width=1pt,-,black] (y) -- (x3);
        \draw [line width=10pt,-,loosely dashed2,black] (x1.south east) to[bend right=40] (x2.south west);
    \end{tikzpicture}
    \caption{\textbf{Choice graph for Debreu's example of the similarity effect.} \\ The vertex $R$ represent the cue node `choose a Record', the other vertices represent the choice alternatives `Beethoven conducted by $F$' $(B_F)$, the equally appealing `Beethoven conducted by $K$' $(B_K)$, and the slightly more appealing `Debussy by the $C$ quartet' $(D_C)$. Whereas the negative relation between $B_F$ and $B_K$ has no influence on choice probabilities for any of the possible two-element subsets, when a choice has to be made from all three alternatives together, it will predict the occurrence of `irrational' yet intuitively expected choice behaviour.}
    \label{fig:02}
\end{figure}

One could argue that the approach of reverse engineering a network structure until one obtains the desired probabilities is a weakness of the framework. We do however believe that this is not a vice but rather a virtue, as it allows the researcher to frame a single process into multiple theoretically distinct structures. And while the expected probability distribution for each of these can be the same, interventions or manipulations will naturally result in distinct predictions. For example, if you were just presented with $\{D_C, B_F, B_K\}$ and chose Beethoven recordings by $K$. If you are then given the opportunity to choose one other recording from the remaining set $\{D_C, B_F\}$, you would probably never choose $B_F$ in this case. This is also what would happen in the current choice structure, as the model now takes the fact that you have $B_K$ already into account ($\sigma_{B_K} = 1$), essentially turning it into an extra cue, such that the negative relation between the two Beethoven recordings makes it unlikely that you will choose the remaining Beethoven recording.   

\subsection*{Attraction}

Whereas the occurrence of the similarity effect has been explained by later models, for example, Tversky's elimination-by-aspects model\cite{tversky1972}, the attraction effect poses a bigger challenge. We consider a simple choice setup from Simonson \& Tversky\cite{simonson1992}, in which one group was offered a reward $(R)$ and had to choose between a nice pen $(P_+)$, which was chosen 36\% of the time, or 6\$ $(\$)$, chosen 64\% of the time. A second group was offered the same choice, now with a second, less attractive, pen $(P_-)$, added to the choice alternatives. As expected the less attractive pen was only chosen 2\% of the time, however the proportion of people choosing the more attractive pen rose to 46\%. 

\begin{figure}
    \centering
    \begin{tikzpicture}[every node/.style = {draw=none, text=black, circle, minimum size = 13mm, fill=gray!25}]
        \path
        (-.5,3) node(3a)[fill=white] {\textbf{a}}
        (3,3) node(y2) {$R$}
        (0,0) node[draw, line width=4.25pt](x21) {$\$$}
        (3,0) node[draw, line width=3.675pt](x22) {$P_+$}
        (6,0) node[draw, line width=.825pt](x23) {$P_-$}
        (8.5,3) node(3b)[fill=white] {\textbf{b}}
        (12,3) node(y) {$R$}
        (9,0) node[draw, line width=1.75pt](x1) {$\$$}
        (12,0) node[draw, line width=1.175pt](x2) {$P_+$}
        (15,0) node[draw, dashed, line width=1.675pt](x3) {$P_-$};
        \draw [line width=1pt,-,black] (y) -- (x1);
        \draw [line width=1pt,-,black] (y) -- (x2);
        \draw [line width=1pt,-,black] (y) -- (x3);
        \draw [line width=1pt,-,black] (y2) -- (x21);
        \draw [line width=1pt,-,black] (y2) -- (x22);
        \draw [line width=1pt,-,black] (y2) -- (x23);
        \draw [line width=2.75pt,-,loosely dashed,black] (x21.south east) to[bend right=40] (x23.south west);
        \draw [line width=.59pt,-,loosely dashed,black] (x21.south east) to[bend right=40] (x22.south west);
        \draw [line width=2.75pt,-,black] (x2.south east) to[bend right=40] (x3.south west);
    \end{tikzpicture}
    \caption{\textbf{Choice graph for Simonson \& Tversky's example of the attraction effect.} \\ The vertex $R$ represent the cue node `choose a Reward', the other vertices represent the choice alternatives `money' $(\$)$, the slightly less appealing `nice pen' $(P_+)$, and the much less appealing `pen' $(P_-)$. \textbf{a}: the attraction affect is explained by imposing negative relationships between the money and each of the pens. \textbf{b}: the attraction effect is explained by imposing a positive relationship between the two pens.}
    \label{fig:03}
\end{figure}

A key assumption of rational choice is that of regularity, meaning that introducing a new option cannot make the choice probabilities of the other alternatives increase. It is clear that assumption this is violated in the current example, as adding a less appealing pen boosted the attraction of the nice pen, thereby increasing its choice probability. Figure \ref{fig:03} shows two possible choice graphs for this experiment, both these structures result in expected response frequencies similar to those found in the experiment, however, each of these explain the occurrence of the attraction effect in a different way. While in Figure \ref{fig:01}a the explanation of the attraction effect rests on the presence of a negative association between the money and both of the pens, in Figure \ref{fig:01}b the effect is explained by a positive association between both of the pens. The framework presented in this paper thus provides two ways in which the mere addition of a less appealing alternative, can boost the selection of otherwise less frequently chosen alternatives. 

As mentioned before, the fact that we can obtain the same results from different structures should be seen as virtue of the model, as it allows us to compare the different theories that underlie each of the different structures. For example, both choice graphs in Figure \ref{fig:03} predict that adding the less attractive pen to the set of available alternatives boosts the selection probability of the nice pen with respect to the money. Now, if we would ask someone, who just chose the nice pen, to make a subsequent choice between the money and the less attractive pen, essentially everyone would choose the money. This is however not the case for the one of the choice graphs in Figure \ref{fig:03}. While in the case of graph \ref{fig:03}a the money would now be chosen more than 90\% of the time, for graph \ref{fig:03}b 33\% of the people would still choose the less attractive pen. This illustrates that while both of these graphs explain our current observations, the new predictions that they make are different, where one of the options is obviously much less plausible. The fact that the new framework allows us to make these kinds of predictions, makes theories in this framework thus falsifiable. 

\subsection*{Compromise}

The third violation, known as the compromise effect, can be demonstrated by another example from Tversky and Simonson\cite{tversky1993}. Here one group of subjects was offered a choice between a medium quality camera, with a regular price $(M)$, or a less expensive, but lower quality camera $(L)$. The second group was given the same choice with a high quality, but expensive, camera added to the set of alternatives $(H)$. In the first group the subjects chose both the low and medium quality/price cameras with equal probability. In the second group the medium camera was still chosen approximately half of the time, and the remainder split among the high and low quality/price cameras. As the medium and low camera's are equally appealing in the first group, and the low and high quality camera are equally appealing in the second group, it is rational to judge the medium and high camera also to be equally attractive. As this clearly is not the case, like the attraction effect, adding a new alternative violates the regularity assumption.

\begin{figure}[ht]
    \centering
    \begin{tikzpicture}[every node/.style = {draw=none, text=black, circle, minimum size = 13mm, fill=gray!25}]
        \path
        (7,3) node(y2) {$C$}
        (4,0) node[draw, line width=7.5pt](x21) {$L$}
        (7,0) node[draw, line width=7.5pt](x22) {$M$}
        (10,0) node[draw, line width=7.5pt](x23) {$H$};
        \draw [line width=1pt,-,black] (y2) -- (x21);
        \draw [line width=1pt,-,black] (y2) -- (x22);
        \draw [line width=1pt,-,black] (y2) -- (x23);
        \draw [line width=15pt,-,dashed,black,dash pattern=on 8pt off 16pt] (x21.south east) to[bend right=40] (x23.south west);
    \end{tikzpicture}
    \caption{\textbf{Choice graph for Tversky \& Simonson's example of the compromise effect.} \\ The vertex $C$ represent the cue node `buy a Camera', the other vertices represent the camera alternatives with their respected levels of quality and prize, `Low' $(L)$, `Medium' $(M)$, and `High' $(H)$. As the negative relation between $L$ and $H$ is plays no role for the two-element subset $\{L, M\}$ in group one, both $L$ and $M$ are chosen with equal probability. However, the addition of $H$, that has a negative relation with $L$, frames $M$ as the compromise, thereby boosting the probability for it being chosen from the $\{L, M, H\}$ set.}
    \label{fig:04}
\end{figure}

A possible explanation for this finding might be that (dis)advantages of the medium camera with respect to the high and low quality cameras are rather small, whereas the (dis)advantages between the high and low quality camera are much larger. As such, adding the high quality/price camera frames the medium quality/price camera as the compromise. As is shown in Figure \ref{fig:04}, our explanation of the compromise effect, once again, rests on the presence of a negative relation between two alternatives. 

\subsection*{Endowment}

Besides implementing bounded rationality and explaining the occurrence of well known violations of rationality, the model also gives a surprising explanation for the endowment effect\cite{thaler1980}. The endowment effect describes the tendency of people to value an object higher if they possess it compared to when they do not. 

To illustrate the endowment effect we consider a variation on the Debreu example, where all three recordings are equally attractive. Suppose you are given one of the Beethoven recordings and are presented with the option to exchange it for the Debussy recording. While someone else is given the Debussy recording with the option to exchange it for the Beethoven recording. As both recordings are equally attractive, the choice axiom predict that you would exchange your current Beethoven recording for the Debussy about half the time. The endowment effect says however that people are unlikely to switch, a prediction that has been experimentally verified\cite{kahneman1990}. 

Explanations offered for this effect are choice-supportive bias, the finding that people tend to attribute more positive features to the object they chose compared to the object they did not choose\cite{mather2000}, as this aligns their attitudes to be consistent with the choice and minimises cognitive dissonance. A second mechanism could be the idea of loss aversion, the tendency that people value preventing the loss of an object more than gaining an object of the same value\cite{kahneman1984}. In both cases this would translate to an increase in base appeal of the alternative as soon as it would come into your possession. As such, while two objects can be equally attractive before a choice is made, as soon as the choice is made the chosen object will have an increase in base attractiveness compared to the non chosen object.
 
With the current model we also obtain a new explanation, that does not depend on changes in the values of the choice problem, but instead ties in to the choice process itself. If we frame the choice as a choice between the Debussy of the Beethoven, having been given the Debussy makes that all conditions of the choice problem are met. That is the choice process starts in a configuration which is equivalent to one where a choice has already been made. As such there is no need to iterate at all, which would explain why people tend to stay more often with their initial choice. 

\section*{Discussion}

This article discusses the key phenomena discovered in discrete choice behaviour, and proposes an elegant model, derived from the Ising model from statistical mechanics, with simple dynamics to explain these. The proposed model explains a number of deviations from rationality, and exhibits key theoretical principles from the field of decision making in a natural way, and allows the construction of new choice situations that can predict new types of irrational behaviour. 

It is remarkable that the theoretically distinct violations of rationality, that are well established within the literature, can be explained by a simple mechanism in the choice model, namely the introduction of a (negative) relation between alternatives and the rule to stop as soon as the choice conditions hold for the first time. One might therefore argue that we are not dealing with three different effects that cause a violation of rationality, but just one mechanism that emerges in multiple situations. 

While we introduced the model here in all its simplicity, it is clear that many extensions and alternations are possible. For example, it might be appropriate to introduce a non-symmetric relation between two alternatives in some situations. Also, the decision process used in this paper was chosen for illustrative purposes, naturally several other Markov processes of selecting, comparing, and discarding choice alternatives are possible, all having the appropriate invariant distribution for the choice probabilities. Depending on the particular choice, some processes may be more suitable than others. 

\begin{methods}

\subsection{Metropolis choice process}
For a choice with $n$ alternatives, $2^n$ different configurations for the vector $\boldsymbol{\sigma}$ exist. We use $\boldsymbol{\sigma}_i$ to denote the $i^{th}$ of the $2^n$ possible states. Now, let $\mathbf{P}$ be a square matrix of order $2^n$, where element $P_{ij}$ contains the probability of transitioning from $\boldsymbol{\sigma}_i$ to $\boldsymbol{\sigma}_j$ in one step of the Metropolis algorithm. $P_{ii}$ contains the probability of staying in the current state, either because the proposal state equals the current state, or the proposal state is rejected. 

\begin{align}
    P_{ij} &= s(\boldsymbol{\sigma}_{j} \mid \boldsymbol{\sigma}_i) p(\boldsymbol{\sigma}_i \rightarrow \boldsymbol{\sigma}_j) \\
    P_{ij} &= s(\boldsymbol{\sigma}_i \mid \boldsymbol{\sigma}_i) + \sum_{j \neq i} s(\boldsymbol{\sigma}_{j} \mid \boldsymbol{\sigma}_i) (1-p(\boldsymbol{\sigma}_i \rightarrow \boldsymbol{\sigma}_j))
     \label{eq:04}
\end{align}

From $\mathbf{P}$ we can obtain the expected rational response probabilities, contained in the stationary distribution, by taking the values of the elements of the first eigenvector of $\mathbf{P}$ that correspond to states in which the choice conditions are met (i.e., $\sigma_c = 1$ and $\sigma_+ = 2$), and dividing these by their sum. We will not go into this approach at length as these probabilities can much more simply be obtained from equation \ref{eq:02}.

To obtain the expected response probabilities, for when the response process stops as soon as the choice conditions $\sigma_c = 1$ and $\sigma_+ = 2$ are met for the first time, we reformulate the Markov chain with transition matrix $\mathbf{P}$ as an absorbing chain with $n$ absorbing states, i.e., those states where only one alternative is active, and $2^n - n$ transient states, i.e., those states where more than one alternative or no alternatives are active. $\mathbf{P}$ has canonical form

\begin{align}
   \mathbf{P} = 
 \begin{bmatrix}
  \mathbf{Q} & \mathbf{R} \\
  \mathbf{0} & \mathbf{I}
 \end{bmatrix} \, ,
     \label{eq:05}
\end{align}

\noindent where $\mathbf{Q}$ contains the transition probabilities from one transient state to another transient state, $\mathbf{R}$ contains the transition probabilities from the transient states to absorbing states and $\mathbf{I}$ is an identity matrix of order $n$ representing that once an absorbing state is entered it cannot be left, and $\mathbf{0}$ is the zero matrix, which represents the transition probabilities from an absorbing state to a transient state (which are all zero). Rearranging $\mathbf{P}$ in its canonical form allows us to derive the expected progression of the Markov chain more easily \cite{diederich2003}.

Let $x \in \{1,2,\dots,n\}$ denote the alternative that is chosen. Furthermore, we use $\mathbf{z} = [z_1, z_2, \dots, z_{2^n}]$ to denote the resting state probabilities, where $z_i$ denotes the probability for the system to have configuration $\boldsymbol{\sigma}_i$ for the alternatives in the resting state distribution. That is, $\mathbf{z}$, contains the probabilities of starting the choice process in some particular configuration $(\boldsymbol{\sigma})$ of the alternatives. This vector of probabilities $\mathbf{z}$ can be divided into the part containing the probabilities of starting in an absorbing state($\mathbf{z}_a$), or transient states ($\mathbf{z}_t$). Finally, we let $s$ denote the number of iterations of the Metropolis algorithm.   

We can now write the (marginal) probability that alternative $x$ is chosen as

\begin{align}
    p(x) = \mathbf{z}_a \, \mathbf{I} + \mathbf{z}_t \, \sum\limits_{s = 0}^{\infty} \, \mathbf{Q}^s \, \mathbf{R}_x \,,
         \label{eq:06}
\end{align}

\noindent and using geometric series to rewrite the infinite sum over $\mathbf{Q}^s$ we can simplify this expression as

\begin{align}
    p(\text{x}) = \mathbf{z}_a \, \mathbf{I_x} + \mathbf{z}_t \, (\mathbf{I} - \mathbf{Q})^{-1} \, \mathbf{R}_x \,.
     \label{eq:07}
\end{align}

As an added bonus we can derive the expected number of Metropolis iterations before alternative $x$ is chosen. While this is beyond the scope of the current paper, one can imagine using the number of iterations as a proxy for the expected (qualitative ordering of) response times. Such predictions can in turn be used to validate/compare different choice structures, or different response processes for the same choice structure. The expected number of Metropolis iterations before particular alternative is chosen is given by

\begin{align}
    E(\text{s} \mid \text{x}) &= \frac{1}{p(\text{x})} \, \left[ \mathbf{z}_t \, \sum\limits_{s = 1}^{\infty} \, s \, \mathbf{Q}^{s-1} \, \mathbf{R}_x \right] \, .
     \label{eq:08}
\end{align}

We can again use geometric series to rewrite the infinite sum over $s \, \mathbf{Q}^{s-1}$ and simplify the expression as

\begin{align}
    E(\text{s} \mid \text{x}) &= \frac{1}{p(\text{x})} \, \left[ \mathbf{z}_t \, (\mathbf{I} - \mathbf{Q})^{-2} \, \mathbf{R}_x \right] \, .
     \label{eq:09}
\end{align}
\newpage
\subsection{Parameter values}
For each of the examples in Figures \ref{fig:02} to \ref{fig:04}, we set $\beta = 1$, and $\varepsilon =0$ to make them easily comparable. Furthermore, in all examples the strength of the relationship between the cue's and the alternatives was set to one $(\omega_{ca} = 1)$. All base appeals, and relations between alternatives were set to zero $(\mu = 0, \omega_{aa} = 0)$ unless otherwise specified. The parameters for which non-default values were used are given in Table \ref{tab:01}.

\subsection{Data availability statement} 
Data sharing not applicable to this article as no datasets were generated or analysed during the current study.

\subsection{Code availability}
Software code used for calculating the expected choice probabilities and expected number of Metropolis iterations was written for the R programming environment\cite{R}, and is available from Github at https://github.com/JoostKruis/CIRM/. The code is dependent on the plyr (https://cran.r-project.org/package=plyr), and qgraph (https://cran.r-project.org/package=qgraph) packages. This repository also contains the setup to run this code for all examples discussed in the paper. 
\end{methods}

\newpage

\bibliographystyle{naturemag}
\bibliography{physicsofchoice}

\newpage

\begin{addendum}
 \item This research was supported by NWO (The Dutch organisation for scientific research); No. 022.005.0 (J.K.), No. CI1-12-S037 (G.M.), No. 451-17-017 (M.M.), No. 451-15-008 (D.M.), No. 314-99-107 (H.M.).
 \item[Author Contributions] J.K. and G.M. wrote the manuscript; J.K. created the analytic model with contributions from G.M., M.M., and M.B.; H.M. gave conceptual advice. All authors discussed the results and implications and commented on the manuscript.
 \item[Competing Interests] The authors declare that they have no competing interests.
 \item[Correspondence] Correspondence and requests for materials should be addressed to Joost Kruis~(email: j.kruis@uva.nl).
\end{addendum}

\newpage

\begin{table}[h]
\centering
\caption{Parameter values used in examples.}
\medskip
\begin{tabular}{cllll}
\hline
Figure: & 2 & 3a & 3b & 4\\
\hline
 & $\mu_{B_F} = 3$ & $\mu_{\$} = 4.25$ & $\mu_{\$} = 1.75$ & $\mu_{C_L} = 7.5$\\
$\boldsymbol{\mu}$ & $\mu_{B_K} = 3$ & $\mu_{P_+} = 3.675$ & $\mu_{P_+} = 1.175$ & $\mu_{C_M} = 7.5$\\
 & $\mu_{D_C} \approx 3.405$ & $\mu_{P_-} = 0.825$ & $\mu_{P_-} = -1.675$ & $\mu_{C_H} = 7.5$\\
\hline
$\boldsymbol{\omega}$ & $\omega_{B_F, B_K} = -10$ & $\omega_{\$, P_+} = -0.59$ & $\omega_{P_+, P_-} = 2.75$ & $\mu_{C_L, C_H} = -15$\\
 &  & $\omega_{\$, P_-} = -2.75$ &  & \\
\hline
\end{tabular}
\label{tab:01}
\end{table}

\end{document}